\documentclass[]{spie}

\usepackage{amsmath,amsfonts,amssymb}
\usepackage{graphicx}
\usepackage[colorlinks=true, allcolors=blue]{hyperref}
\usepackage{aas_macros}
\usepackage[version=4]{mhchem}

\title{NEWS: the near-infrared Echelle for wideband spectroscopy}

\author[a]{Mark J. Veyette}
\author[a]{Philip S. Muirhead}
\author[a]{Zachary J. Hall}
\author[a,b]{Brian Taylor}
\author[a,c]{Jimmy Ye}
\affil[a]{Department of Astronomy, Boston University, 725 Commonwealth Ave., Boston, MA, USA}
\affil[b]{TI Research, 2915 Kletha Trail, Flagstaff, Az, USA}
\affil[c]{Department of Physics, College of the Holy Cross, 1 College Street, Worcester, MA, USA}

\authorinfo{Send correspondence to M.J.V., e-mail: mveyette@bu.edu}

\newcommand\afe{[$\alpha$/Fe]}
\newcommand\al{$\alpha$}
 
\begin{document} 
\maketitle

\begin{abstract}
We present an updated optical and mechanical design of NEWS: the Near-infrared Echelle for Wide-band Spectroscopy (formerly called HiJaK: the High-resolution J, H and K spectrometer), a compact, high-resolution, near-infrared spectrometer for 5-meter class telescopes. NEWS provides a spectral resolution of 60,000 and covers the full 0.8--2.5 $\mu$m range in 5 modes. We adopt a compact, lightweight, monolithic design and developed NEWS to be mounted to the instrument cube at the Cassegrain focus of the the new 4.3-meter Discovery Channel Telescope.
\end{abstract}

\keywords{infrared astronomy, infrared spectroscopy, low-mass stars, exoplanets, stellar abundances}
	
\section{Introduction}
\label{sec:intro}

High-resolution, near-infrared (NIR) spectroscopy enables an enormously broad range of scientific studies (see Ref.~\citenum{Muirhead2014} and references therein). However, relatively few facility-class, high-resolution, NIR spectrometers currently exist. Most are only available on large, heavily subscribed 8-to-10 meter class telescopes, such as NIRSPEC on the 10-meter Keck II Telescope\cite{McLean1998}, CRIRES on the 8.2-meter VLT UT 1 Telescope\cite{Kaeufl2004}, and IRCS on the 8.2-meter Subaru Telescope\cite{Tokunaga1998,Kobayashi2000}. High-resolution, NIR spectrometers for 3--5-meter class telescopes like Lowell Observatory's new 4.3-meter Discovery Channel Telescope\cite{Levine2012} (DCT) in Happy Jack, Arizona would provide greater accessibility for this powerful yet under-utilized tool for astronomy.

Offering continuous, wide-band coverage has been an obstacle for high-resolution, NIR spectrometers. The number of resolution elements ($\Delta\lambda$) within the free spectral range (FSR) of a single order of a grating-based spectrometer is given by
\begin{equation}
    \frac{\mathrm{FSR}}{\Delta\lambda} = \frac{\lambda N}{\phi D},
\end{equation}
where $\lambda$ is the wavelength at the center of the order, $N$ is the number of illuminated grooves, $\phi$ is the angular width of the slit, and $D$ is the diameter of the telescope. Traditionally, higher resolution is achieved by increasing $N$ by using a larger grating or a grating with more closely spaced grooves. In the IR, the number of resolution elements per order quickly becomes too large to fit a full order across a single detector with at least two pixels per resolution element to fully sample the spectrum. 

Immersion gratings provide one path to high resolution while maintaining small free spectral ranges. Spectral resolution increases linearly with the index of refraction of the medium the grating is immersed in. IGRINS\cite{Yuk2010} and iSHELL\cite{Rayner2012} both achieve high-resolution ($R >$ 40,000) across the NIR through the use of a silicon ($n=3.4$) immersion grating. However, silicon is not transmissive below 1.2 $\mu$m. At constant resolution, the 0.8 to 1.2 $\mu$m range accounts for over 35\% of the information content in the 0.8 to 2.5 $\mu$m range. As we discuss in Section~\ref{sec:science}, Y-band around 1 $\mu$m is a requirement for our primary science goal.

Resolution also increases linearly with $\tan(\delta)$, the tangent of the blaze angle. With a high-blaze Echelle grating, high resolution can be achieved in a format that can be imaged by a single 2k$\times$2k detector. Here we present an optical design for a high-resolution NIR spectrograph called NEWS: the Near-infrared Echelle for Wide-band Spectroscopy. The design is based on a high-blaze R6 ($\tan(\delta) = 6$) Echelle grating and achieves a resolution of 60,000 over the full 0.8--2.5 $\mu$m range. The photometric z, Y, J, H, and K bands can be observed in their entirety without gaps.

\section{Scientific Motivation}
\label{sec:science}

The primary science goal of NEWS is to measure the abundances of individual elements in M dwarf stars. One application of this capability is to measure the ages of field M dwarfs. M dwarf stars are the most common class of star in the Galaxy, accounting for 70\% of all stars \cite{Bochanski2010}. Their small radii and low effective temperatures ($2500 \mathrm{K} < T_\mathrm{Eff} < 3800 \mathrm{K}$) make nearby M dwarfs well suited for the detection of small, potentially habitable planets. Results from NASA's \textit{Kepler} Mission suggest M dwarfs are teeming with planets with $\sim$1 rocky planet per M dwarf with a period $<$150 days \cite{Morton2014,Dressing2015}. NASA's \textit{Transiting Exoplanet Survey Satellite} (\textit{TESS}) is expected to discover over 400 Earth-sized planets around nearby M dwarfs, including $\sim$50 orbiting within their host stars' habitable zones \cite{Sullivan2015}.

M dwarfs' low effective temperatures make them difficult to characterize due to the formation of molecules throughout their atmospheres. Their visible and NIR spectra are dominated by millions of molecular lines that blend together even at high resolution. These molecular features render useless the standard methods developed for inferring fundamental parameters like $T_\mathrm{Eff}$, surface gravity, and chemical composition of Sun-like stars.

The age of an M dwarf is perhaps the most challenging fundamental parameter to measure but enables a wide variety of stellar and exoplanet science. M dwarfs are known to host a wide range of planetary architectures: single short-period gaseous planets (e.g. GJ 1214 b \cite{Charbonneau2009}), compact multiple systems (e.g. \textit{Kepler}-42, \textit{Kepler}-445, and \textit{Kepler}-446 \cite{Muirhead2012a,Muirhead2015}), single rocky planets, and multi-planet systems with a wide range of planet masses. Ages of the host M dwarfs in these systems would answer outstanding questions on planet formation and evolution. It is an open question whether the scarcity of short-period gaseous planets is due to lower disk surface density around M dwarfs \cite{Dressing2013} or if they are a short-lived evolutionary state, evaporated by high UV flux \cite{Lopez2014}. Theorists disagree on the timescales for orbital evolution of short-period planets orbiting M dwarfs. Ref.~\citenum{Wu2002} argue the tidally-induced eccentricity-damping timescale for short-period, low-mass planets is small, such that they should be circularized by 1 Gyr. However, Ref.~\citenum{Jackson2008} argue that eccentricity-damping is coupled to semi-major axis damping, extending the timescale to many billions of years. These competing hypotheses are testable given a means to measure M dwarf ages.

Unlike solar-type stars, main-sequence M dwarfs move imperceptibly on a color-magnitude diagram. Gyrochronology, or the study of stellar spin-down versus age, holds some promise for measuring M dwarf ages. However, recent studies of M dwarfs with age-dated white dwarf companions suggest that M dwarfs do not spin down efficiently \cite{Morgan2012}, and can hold onto their rapid rotation for billions of years \cite{West2015,Newton2016}.

\begin{figure}
    \begin{center}
    \includegraphics[width=3.5in]{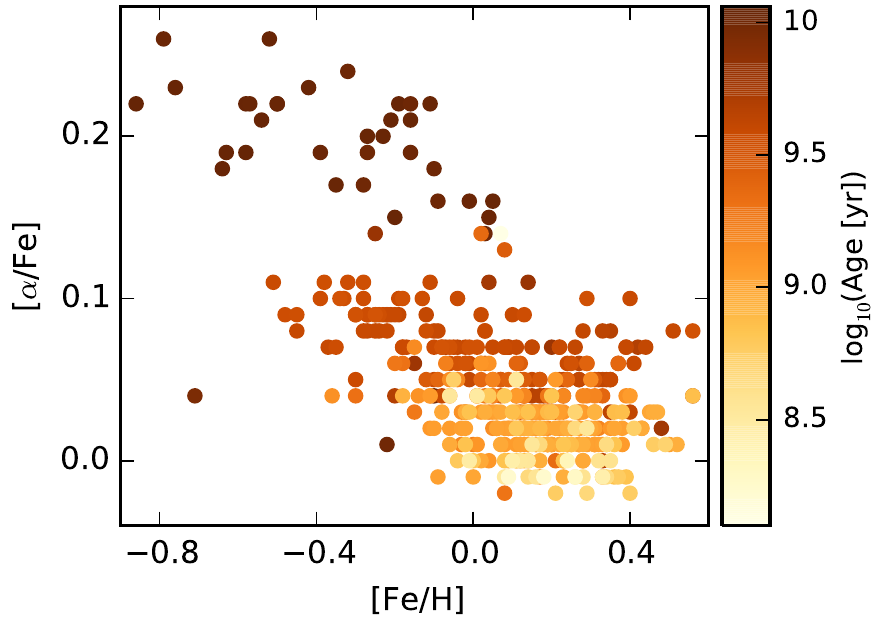}
    \end{center}
    \caption[example]
    {\label{fig:afe_feh_age}
    The \afe{}-[Fe/H]-age relation based on nearby red giants (data from Ref.~\citenum{Feuillet2016}). This relation can provide ages for field M dwarfs given a method to measure \afe{} in M dwarfs.}
\end{figure}

Our knowledge of the chemical evolution of the Galaxy provides a novel tool for estimating the ages of field M dwarfs. Early in the life of the Galaxy, core-collapse supernovae enrich the interstellar medium with the \al{}-elements O, Ne, Mg, Si, S, Ca, and Ti. Subsequently, type Ia supernovae contribute large amounts of Fe and the relative abundance of \al{}-elements to Fe (\afe{}) decreases. Surveys of nearby Sun-like stars and red giants find an empirical relation between \afe{}, [Fe/H], and age \cite{Haywood2013, Bensby2014, Feuillet2016}. Figure~\ref{fig:afe_feh_age} shows the \afe{}-[Fe/H]-age relation for nearby red giants. We determined \afe{} and [Fe/H] can be used to measure ages to an accuracy of $\pm1$ Gyr root-mean-square. This relation provides a new, powerful tool for estimating ages of field M dwarfs, given methods to measure their \afe{} and [Fe/H]. The correlation between \afe{}, [Fe/H], and age is not perfect given the chaotic nature of star formation and availability of pristine gas even in the late stages of galaxy evolution. Ref.~\citenum{Martig2015} found that 6\% of nearby red giants with \afe{} $>$ 0.13 are younger than 6 Gyr. Nevertheless, measurements of \afe{} and [Fe/H] can be combined into a powerful statistical tool for estimating ages of field M dwarfs.

Recently multiple methods for measuring metallicity of M dwarfs have been empirically calibrated via widely-separated binary systems composed of an M dwarf with an FGK companion. The two stars are assumed to have formed at the same time, from the same material and, therefore, share a common chemical composition. Metal-sensitive indicators in M dwarf spectra can be calibrated on metallicities measured from the FGK companion. Methods have been developed to measure M dwarf metallicity from high-resolution NIR spectra \cite{Onehag2012,Lindgren2016}, high-resolution optical spectra \cite{Pineda2013,Neves2014,Maldonado2015}, moderate-resolution NIR spectra \cite{Rojas2010,Rojas2012,Terrien2012,Mann2013a,Newton2014}, and optical-NIR photometry \cite{Bonfils2005,Casagrande2008,Johnson2009,Schlaufman2010,Neves2012,Johnson2012,Hejazi2015}.

\begin{figure}
    \begin{center}
    \includegraphics[width=\linewidth]{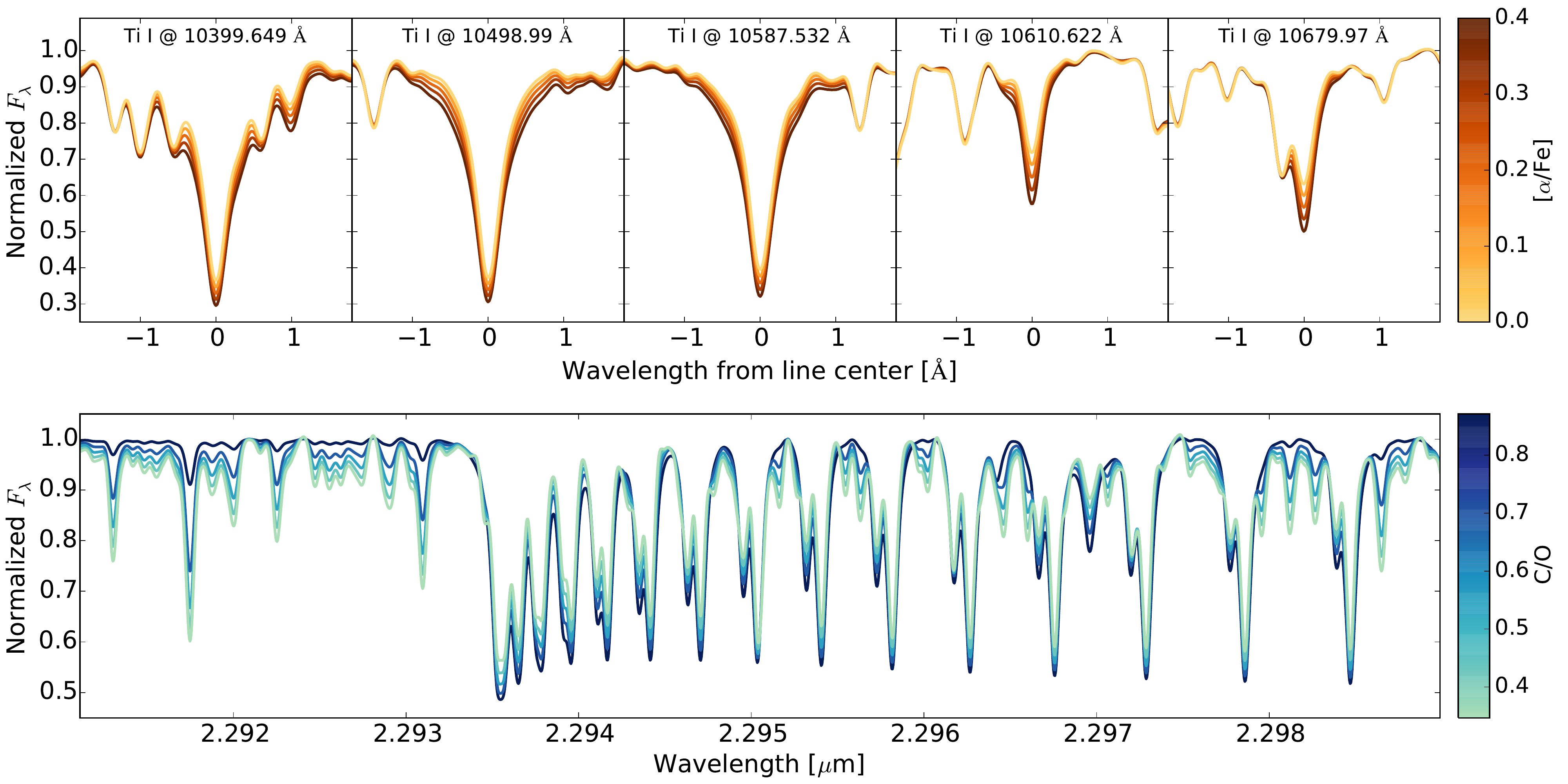}
    \end{center}
    \caption[example]
    {\label{fig:alphalines}
    Psuedo-continuum normalized synthetic M dwarf spectra ($T_\mathrm{eff}$=3000 K, log($g$)=5.0, [M/H]=0.0) smoothed to $R$ = 60,000. Top: A representative sample of deep, \afe{}-sensitive \ce{^{48}}\ion{Ti}{1} lines in Y-band for models with varied \afe{}. Bottom: K-band metal lines and the CO 2-0 bandhead for models with varied C/O. High-resolution Y-band and K-band observations can be used to measure \al{}-abundance in M dwarf stars and calibrate out any effects of C and O abundance.}
\end{figure}

Precise methods for measuring both \afe{} and [Fe/H] are needed to estimate ages of field M dwarfs. Currently, no method exists to measure \al{}-enhancement. Based on \texttt{PHOENIX} BT-Settl synthetic spectra, we found that Y-band around 1 $\mu$m contains numerous \al{}-sensitive lines including many deep \ce{^{48}}\ion{Ti}{1} lines. \ce{^{48}Ti} is a rapid decay product of \ce{^{48}Cr}, a product of the \al{} process during core collapse supernovae. Figure~\ref{fig:alphalines} shows synthetic spectra of M dwarfs with varied abundance of the major \al{} tracers Mg, Si, and Ti. Y-band contains numerous isolated \ce{^{48}}\ion{Ti}{1} lines whose strength correlate with \afe{}. However, these atomic lines are embedded within molecular absorption bands of TiO and FeH. High spectral resolution is required to sufficiently isolate atomic \ce{^{48}Ti} lines to measure \afe{} in M dwarfs. 

It has recently been discovered that the relative abundances of carbon and oxygen strongly affect the psuedo-continuum level throughout M dwarf spectra \cite{Veyette2016}. In order to accurately measure \afe{} in M dwarfs, we must first calibrate out any effect of C and O abundances (or their ratio\footnote{C/O is defined as $N_\mathrm{C}/N_\mathrm{O}$, where $N_\mathrm{C}$ and $N_\mathrm{O}$ are the number densities of carbon and oxygen, respectively.} C/O) on the equivalent widths (EWs) of \ce{^{48}}\ion{Ti}{1} lines in Y-band. Ref.~\citenum{Tsuji2014,Tsuji2015,Tsuji2016} found that high-resolution observations of \ce{H2O} and CO lines in K-band can be used measure C and O abundances in M dwarfs. Figure~\ref{fig:alphalines} shows synthetic spectra of M dwarfs with varied C/O. In the atmosphere of an M dwarf, nearly all the C is locked away in energetically favorable CO, along with an equal amount of O. At $T_\mathrm{eff} < 3300$ K, the majority of the remaining O is found in \ce{H2O}. The strength of the 2.3 $\mu$m CO 2-0 bandhead and the 1.9 $\mu$m \ce{H2O} band can be used to measure C/O.

The science case above led us to the following design requirements. NEWS must achieve high-resolution ($R >$ 30,000) and cover a large portion of the NIR window, from Y-band to K-band. We find that a resolution of $R$ = 60,000 provides a good balance between isolating atomic lines in M dwarf spectra but still allowing broadband observations in a single exposure. The design must also offer high throughput ($>$ 10\%) in order to achieve high enough signal-to-noise to measure small changes in the EWs of atomic lines in M dwarf spectra. This requirement led us to use a slit-fed design mounted directly to the telescope as opposed to a bench-mounted, fiber-fed design which can be limited by modal noise (e.g. GIANO\cite{Origlia2014}). Although the ability to measure individual elemental abundances and ages of M dwarfs has dictated many of the design requirements for NEWS, we adopted an overall facility-class philosophy. Specifically, we designed NEWS to cover the full 0.8--2.5 $\mu$m range without gaps and offer multiple slit widths and lengths. We also developed NEWS to be extremely compact and lightweight so that the design can be implemented at nearly any 3--5 meter class telescope.

\section{Optical Design}
\label{optics}

\subsection{From Telescope to Detector}
\begin{figure}
	\begin{center}
	\includegraphics[width=3.5in]{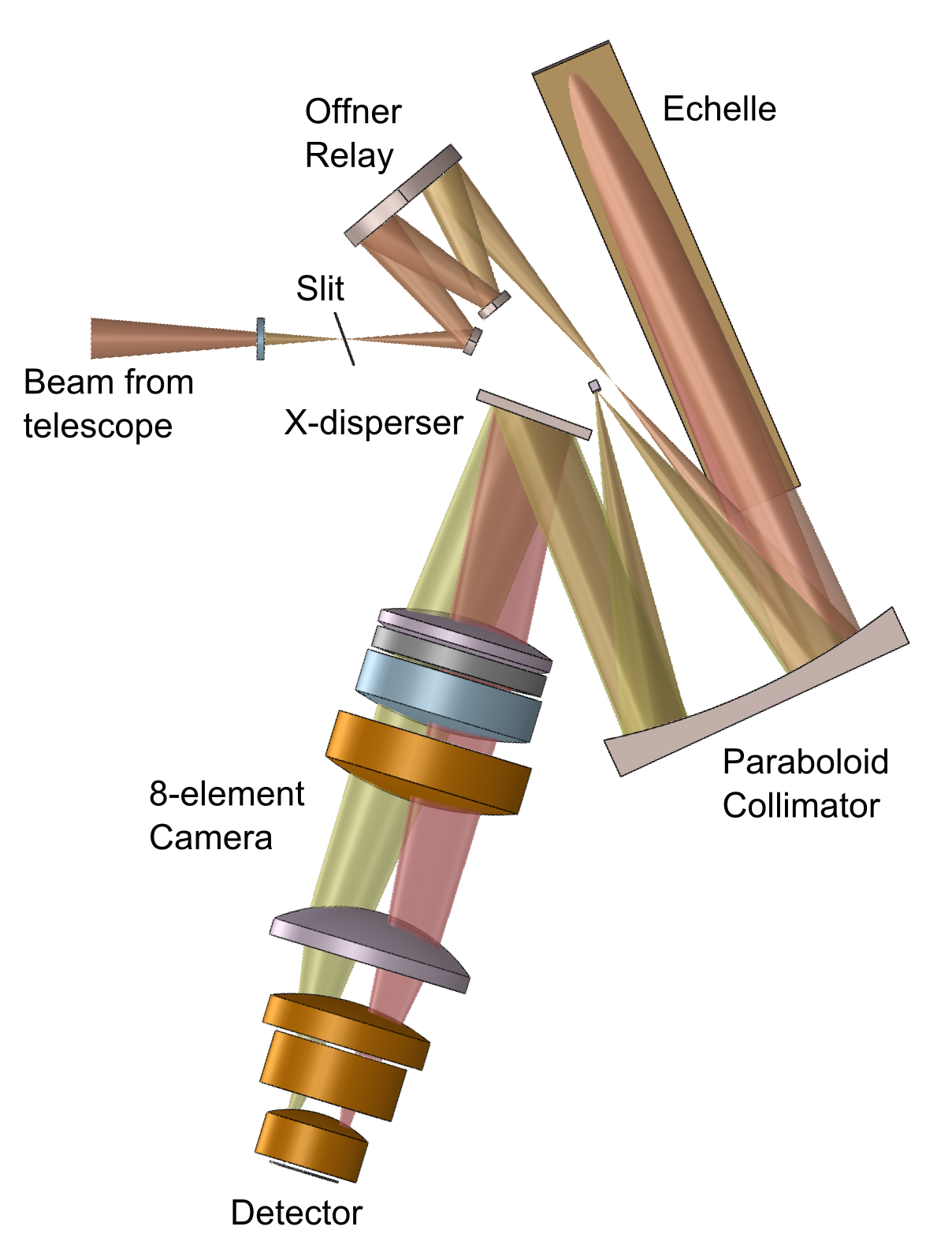}
	\end{center}
	\caption[example] 
	{\label{fig:optics} 
	 CAD rendering of the NEWS optical layout.}
\end{figure}

We show the full NEWS optical layout in Figure~\ref{fig:optics} and list the key properties of the design in Table~\ref{tab:news}. The converging f/6.1 beam delivered from the telescope enters the cryostat through a fused silica window. The beam comes to a focus at one of six slits accessible by a slit wheel before passing through one of five order-selecting filters accessible by a filter wheel. The beam is then folded onto an Offner relay that serves as a cold pupil stop. After exiting the Offner relay, the beam goes through a second focus and expands to a diameter of 4 cm before being collimated by a parabolic mirror used off-axis. The collimated beam is reflected by an R6 Echelle grating that is tilted from Littrow by an in-plane angle of $\theta$ = 2\degr{} and by an off-plane angle of $\gamma$ = 1.5\degr. The dispersed beam is refocused by the same parabolic mirror and is folded by a Mangin mirror to remove aberrations introduced by dispersing onto the parabolic mirror and direct the beam to the conjugate point of the same parabolic mirror. The re-collimated beam is cross-dispersed by a grating before being focused by an 8-element, all-spherical camera onto a Hawaii 2-RG detector.

\subsubsection{Slit Selection}
The NEWS design equips a motorized wheel to cycle through six different slit options. The width options for the slits are either 0\farcs{}5, 1\farcs{}0, or 1\farcs{}5 in the dispersion direction to enable observations during nights with poor seeing at reduced resolution.  The silts are either 5\arcsec{} or 9\arcsec{} long. In order to fully separate adjacent orders, only the 5\arcsec{} slit can be used with the 6th and 7th orders of the cross-disperser (corresponding to z- and Y-band). Slits will be laser-cut into 2-inch diameter gold-coated silicon wafers. Tilted slit substrates direct a $\sim$6 arcmin$^2$ field back out of the cryostat where a slit-viewing camera images it onto an InGaAs detector.

\subsubsection{Collimation}
A monolithic parabolic mirror collimates the beam for both the Echelle and cross-disperser. The NEWS design employs a Mangin mirror to remove aberrations introduced by dispersing onto the parabolic mirror and to steer the beam returned by the Echelle to the conjugate point of the paraboloid for collimation onto the cross-disperser. The Mangin reflector is a spherical \ce{CaF2} lens with curvature on both sides and a reflective coating on one side.

\subsubsection{Primary Dispersion}
We designed NEWS around a new R6 Echelle recently developed by Richardson Gratings with a coarse groove spacing and high blaze angle (13.3 grooves/mm, blazed at 80.5\degr). The high blaze angle and coarse grooves enable high spectral resolution with a manageable number of resolution elements per order so that the entire echellogram can be imaged by a single square detector with sufficient sampling. The Echelle is tilted in-plane by $\theta$ = 2\degr{} in order return the full 0.8--2.5 $\mu$m range without gaps. The in-plane tilt reduces the peak efficiency by $\sim$30\%. Th Echelle is also titled off-plane by $\gamma$ = 1.5\degr{} to separate the incoming and outgoing beams at the intermediate focus.

\subsubsection{Cross Dispersion}
For cross-dispersion, we utilize the same grating that is used as the primary dispersive grating in TripleSpec\cite{Herter2008} and GNIRS\cite{Elias1998}. The grating has 110.5 groves/mm and is blazed at 22\degr. Orders 7, 6, 5, 4, and 3 of the cross-disperser correspond roughly to the z, Y, J, H, and K atmospheric windows, respectively. Only a single cross-dispersion order is accessible at a time and is selected by order-blocking filters.

\subsubsection{Spectrograph Camera}
\begin{figure}
	\begin{center}
	\includegraphics[width=\linewidth]{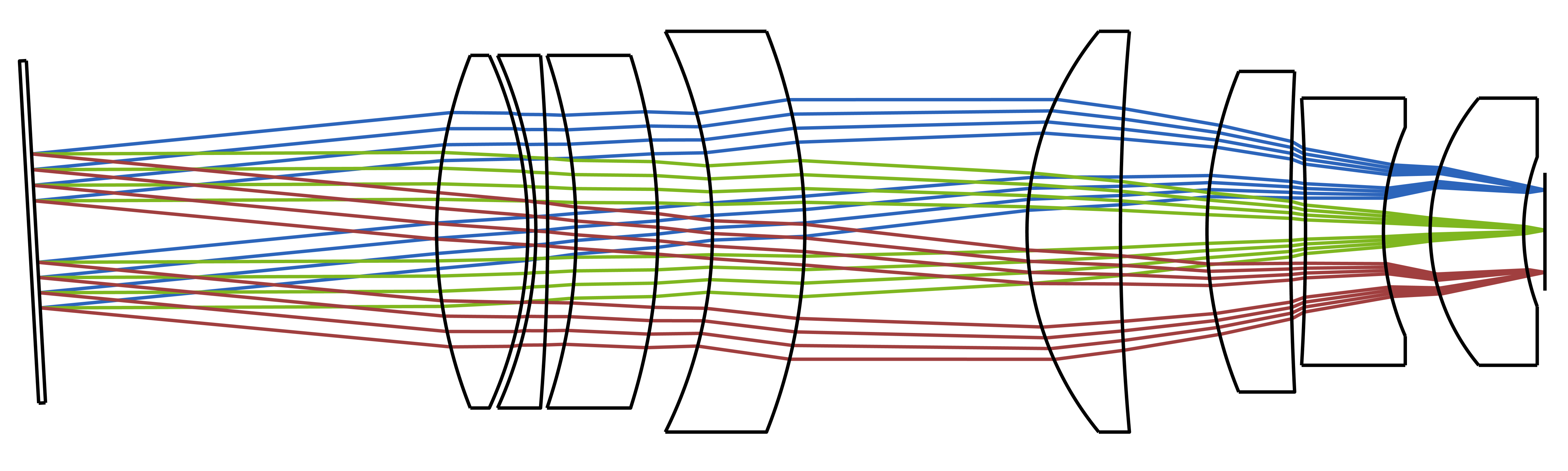}
	\end{center}
	\caption[example] 
	{\label{fig:camera} 
	 Camera Optical layout.}
\end{figure}

We designed an 8-element, all-spherical f/2.6 camera to image the spectrum onto a 2k$\times$2k Teledyne Hawaii 2-RG detector with a 2.5 $\mu$m cutoff. The camera consists of two \ce{CaF2}, one Infrasil, one Fused Silica, and 4 ZnSe lenses. Figure~\ref{fig:camera} shows the optical layout of the camera.

\begin{table}
\centering
\small 
\begin{tabular}{ | p{1.8in} | p{=4.25in} |  }
 \hline
 \multicolumn{2}{|c|}{\bf NEWS Fundamental Properties} \\
 \hline
Resolution ($\lambda / \Delta \lambda$) & 60,000 for 0\farcs{}5 slit width, 30,000 for 1\farcs{}0 slit width\\

Wavelength coverage & 5 bands: Either 0.80-1.0 ($z$), 1.00-1.20 ($Y$), 1.20-1.45 ($J$), 1.45-1.85 ($H$) or 1.85-2.5 $\mu$m ($K$), selectable by filter wheel\\

Slit sizes on sky & 6 unique sizes: either 0\farcs{}5, 1\farcs{}0 or 1\farcs{}5 wide (dispersion directions), and either 9\farcs{}0 or 5\farcs{}0 long, laser cut into Au-coated Si substrates, selectable by slit wheel. Optimal slit length choice depends on band due to x-dispersion.  One substrate lacks a slit for acquiring cold darks.\\

Beam size on Echelle & 1.57 inches (4 cm)\\

End-to-end throughput & 5\% requirement, 10\% goal\\

DCT Properties & 4.3-meter diameter, Alt/az, Ritchey–Chretien, f/6.1 at Cassegrain, 127 $\mu$m per arcsecond\\

Field Rotation & Instrument mount rotation built into DCT Cassegrain cube\cite{Bida2012}\\

Atm. Disp. Correction & None (slit can be rotated to parallactic angle via the Cassegrain cube)\\

Slit Viewer Camera & COTS InGaAs detector and short-wave infrared (SWIR) lens system external to cryostat operated in $J$ band (similar to NIHTS\cite{Bida2014})\\

Guiding & Native DCT Off-axis guider\\

Cold stop & Offner relay (identical to NIHTS\cite{Bida2014})\\

Echelle & Newport/Richardson 53-*-182E operated in quasi-Littrow (13.33 l/mm, 80.54\degr Blaze)\\

Cross-disperser & Newport/Richardson 53-*-138R operated with in-plane tilt (110.5 l/mm, 22\degr Blaze, identical to TSPEC\cite{Herter2008} and GNIRS\cite{Elias1998})\\

Spec. Camera & 8-element, all spherical, f/2.6, \ce{CaF2}, ZnSe, Infrasil, and fused silica lenses\\

Spec. Detector & 2.5-$\mu$m-cutoff Teledyne Hawaii-2RG (2048x2048 18-$\mu$m pixels)\\

Spec. Detector Electronics & Teledyne SIDECAR or Leach and custom detector interface board \\

Sampling & 2.0 pixels per 0\farcs{}5 resolution element (at R=60,000)\\

Cryostat & Box-style, aluminium (e.g. Atlas Ultrahigh Vacuum)\\

Optical Bench & 39\,$\times$\,25\,$\times$\,12 in (99\,$\times$\,64\,$\times$\,30 cm) custom honey-combed aluminum\\

Optical Bench Temperature & Less than 100 Kelvin via a CTI-1050 cryodyne refrigeration system, 65 W max load, similar design to Mimir \cite{Clemens2007}\\

Detector Temperature & Less than 78 Kelvin via a CTI-1050 cryodyne refrigeration system, similar design to Mimir \cite{Clemens2007}\\

Total weight & $<$200 kg, within limit for the Cassegrain cube \\

\hline
\end{tabular}
\caption{News Fundamental Parameters}
\label{tab:news}
\end{table}

\subsection{Optical Performance}
\label{perf}
\begin{figure}
	\begin{center}
	\includegraphics[width=\linewidth]{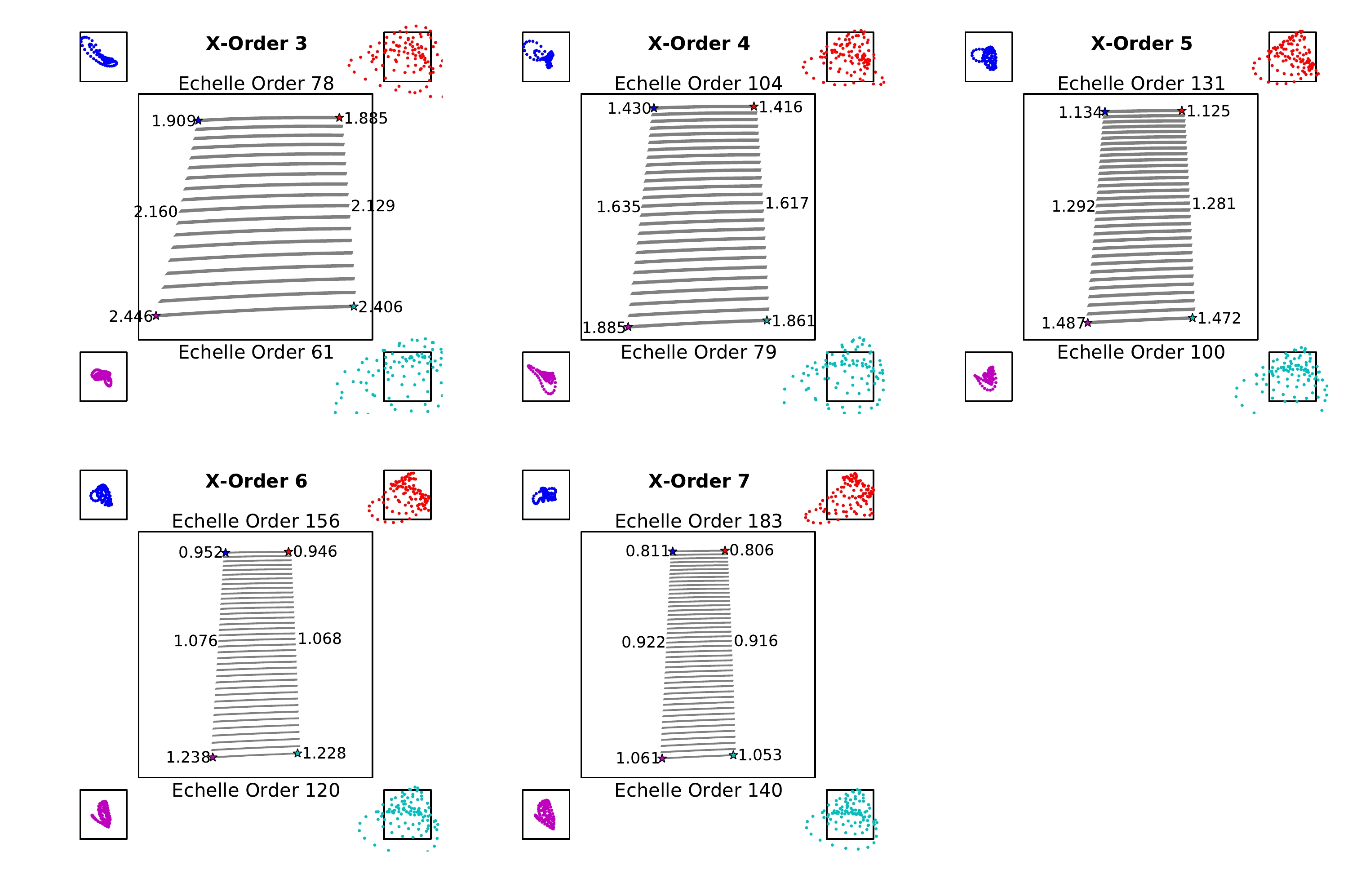}
	\end{center}
	\caption[example] 
	{\label{fig:echellogram} 
	 Ray-traced echellograms and corner spot diagrams for each observing mode of NEWS, selectable entirely by filter wheel. The filter selects an order of the cross-dispersing grating (orders 3 through 7), resulting in wavelength coverage from 0.8 to 2.5 $\mu$m. The large boxes represent the size of the full Hawaii-2RG detector (36.9 x 36.9 mm) and the small boxes represent the size of each pixel (18 x 18 $\mu$m). By using a Mangin reflector in combination with a conjugate paraboloid, aberrations introduced by the collimating paraboloid are reduced to roughly a pixel. The aberrations from the paraboloid dominate the RMS spot sizes so that the tolerances on the optical surfaces are large (typically 10 $\mu$m), except for the Offner relay (1 $\mu$m).}
\end{figure}

We optimized the camera lenses and Mangin fold mirror to minimize the RMS spot size across the detector. The largest source of aberration is wavelength-dependent field curvature introduced by dispersing onto the paraboloid, much of which is removed by the Mangin mirror. As a result, tolerances on the optical surfaces are quite large ($\sim$ 10 $\mu$m). In Figure~\ref{fig:echellogram}, we show the full simulated echellogram and corner spot diagrams. RMS spot sizes are on the order of one 18 $\mu$m square pixel.

\section{Mechanical Design}
\label{mech}

\begin{figure}
	\begin{center}
	\includegraphics[width=4in]{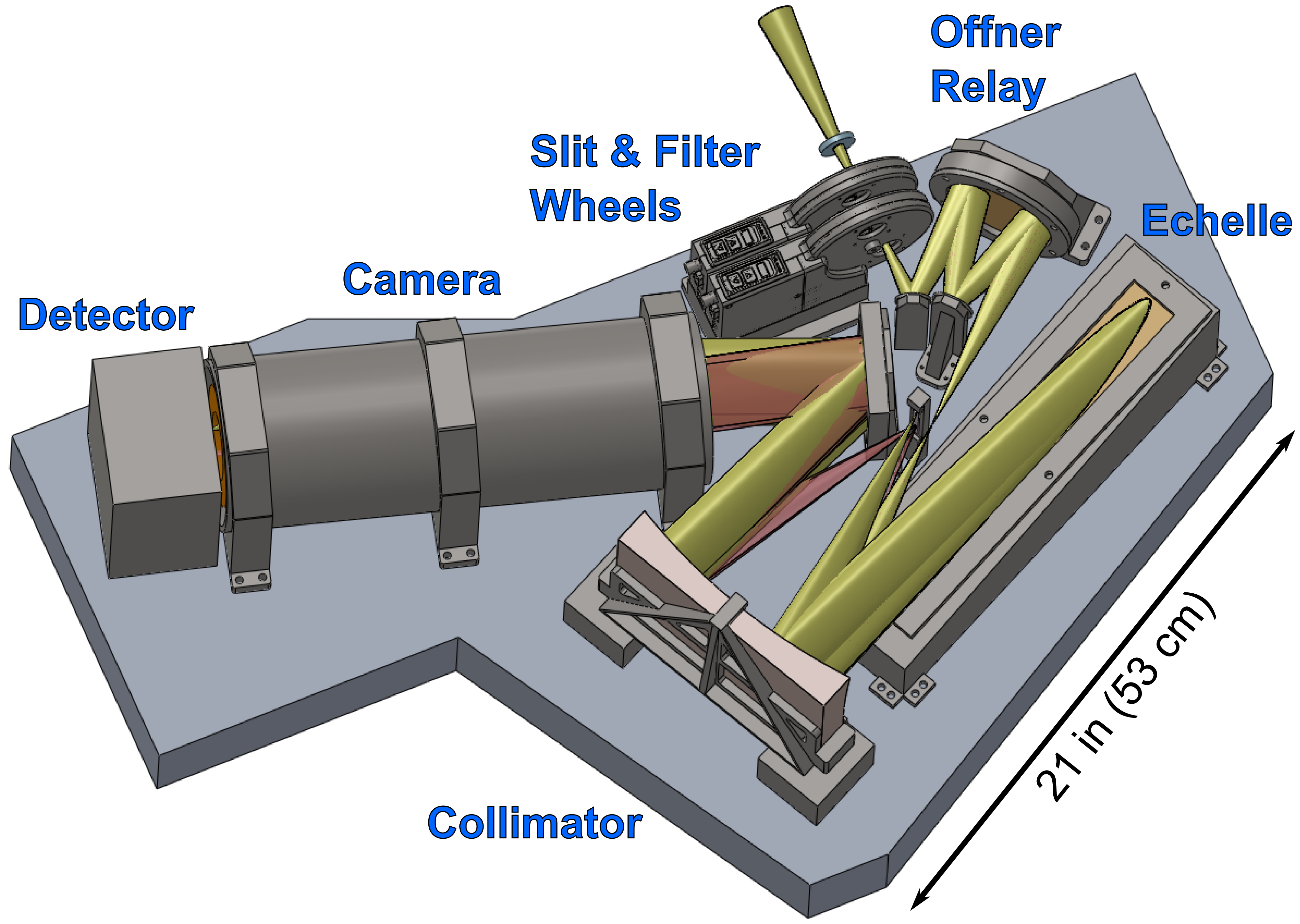}
	\end{center}
	\caption[example] 
	{\label{fig:mech} 
	CAD rendering of the full NEWS optomechanical design. The optical bench and all mounts will be custom machined from 6061-T6 Al.}
\end{figure}

\subsection{Optomechanical design}
Figure~\ref{fig:mech} shows the full NEWS optomechanical design.

\subsubsection{Optical bench}
\begin{figure}
	\begin{center}
	\includegraphics[width=4in]{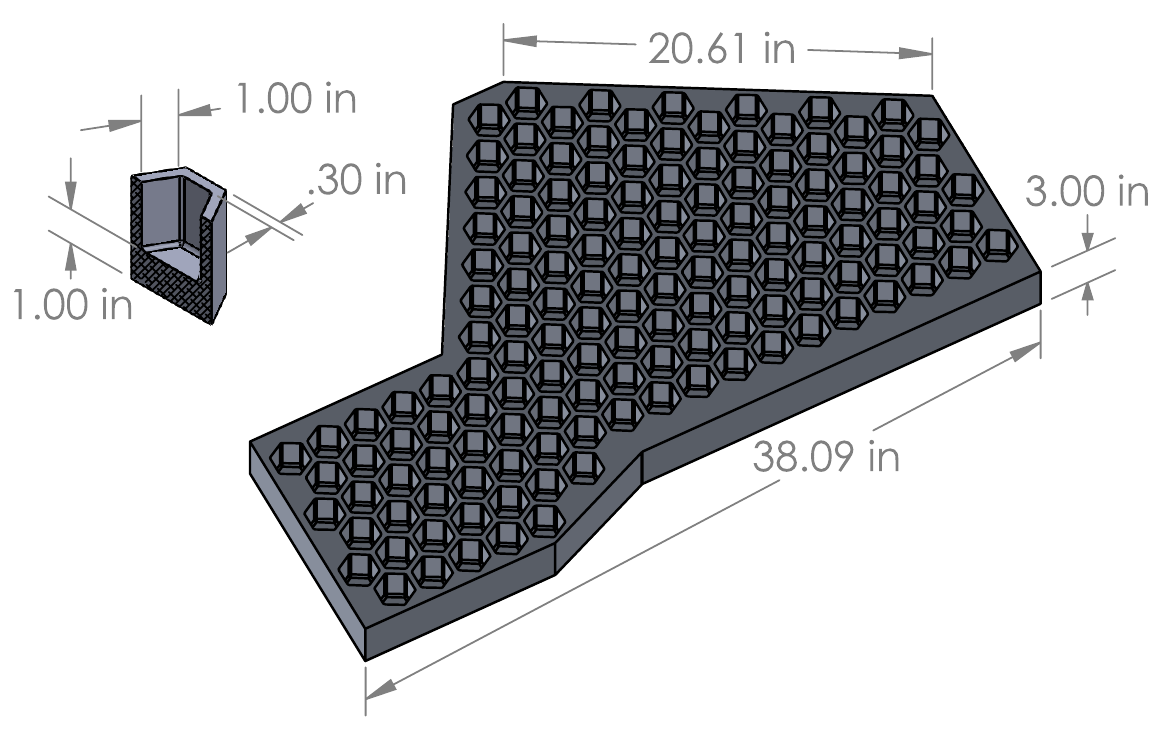}
	\end{center}
	\caption[example] 
	{\label{fig:bench} 
	CAD rendering of the NEWS custom honeycomb optical bench (the bottom side as shown in Figure~\ref{fig:mech}). An inset on the left shows a cross section of one of the honeycomb hexagons. The hexagons are 1 inch on a side (inner) with 0.3 inch thick walls and are milled 2 inches deep, leaving 1 inch of solid aluminum for the mounting surface.}
\end{figure}

To simplify mounting and maintain a compact design, we designed all the optical components of NEWS to lie in a single plane. The mounts for each optical component all attach directly to a monolithic optical bench milled from a single block of 6061-T6 aluminum. To minimize weight, but maintain rigidity, the optical bench is light-weighted with a honeycomb structure. The hexagons of the honeycomb are milled 2 inches deep with 1 inch sides (inner), leaving 0.3 inch thick walls and 1 inch of solid aluminum for the mounting surface. Figure~\ref{fig:bench} shows the light-weighted honeycomb structure. The optical bench will be surrounded by a low-emissivity radiation shield and mounted to an enclosing cryostat via G10 tabs.

We applied finite element analysis (FEA) on the optical bench to ensure it meets mechanical tolerances ($<$ 10$\mu$m flex) under under the weight of the optical elements subject to varied gravity vectors.

\subsubsection{Optics Mounts}
\begin{figure}
	\begin{center}
	\includegraphics[width=2in]{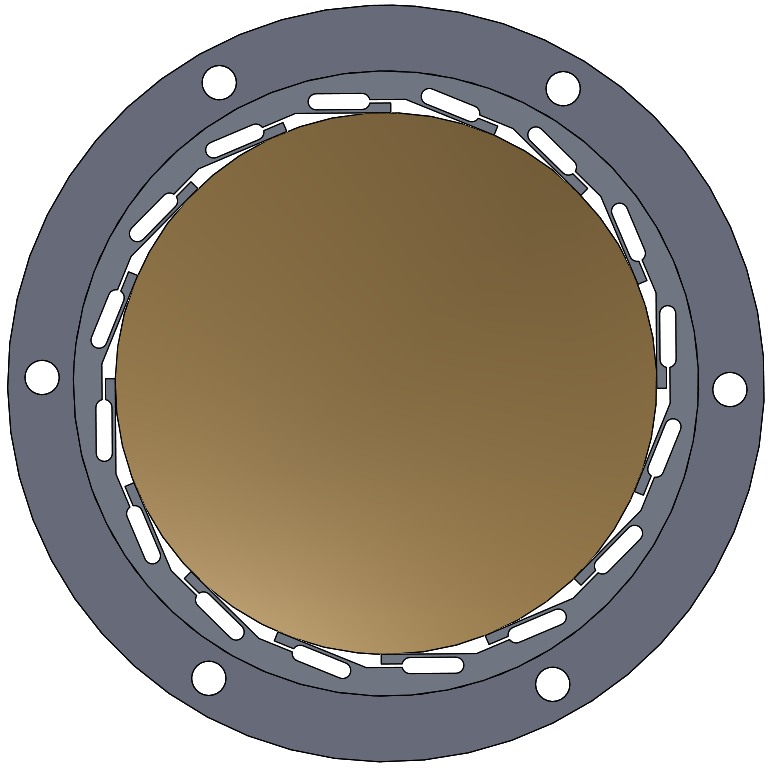}
	\end{center}
	\caption[example] 
	{\label{fig:flexure} 
	CAD rendering of radial flexure ring mount for the Offner relay primary mirror. Wire-cut spring restraints compensate for thermal contraction stress and preload against a changing gravity vector.}
\end{figure}

We designed all optical mounts to be machined out of 6061-T6 Al. To maintain optical positioning tolerances, we employ flexure supports on all optics to compensate for thermal contraction stress and to preload against a changing gravity vector. For example, the Offner relay primary mirror is constrained by a radial flexure ring with 16 EDM wire-cut spring restraints (see Figure~\ref{fig:flexure}). The design is identical to that used in NIHTS\cite{Bida2014}.

\subsection{Mounting to the DCT}
We designed NEWS within the size and weight restrictions of mounting to one of the large ports on the DCT's instrument cube at the Cassegrain focus of the telescope. The instrument cube rotates to maintain constant field alignment. Instruments mounted to the cube must clear the telescope mount supports at all rotation angles and the mount platform at all altitudes. Mounting directly to the telescope entails strict weight constraints. The total payload capability for the instrument cube is 1500 kg\cite{Smith2010b}. The large instrument ports each can support up to 360 kg\cite{Bida2012}. Our current design fits within a 99\,$\times$\,64\,$\times$\,30 cm in volume and weighs less than 100 kg (excluding the cryostat).

\subsubsection{Software System}
\label{soft}
Software control and interfacing to the TCS will be done with the Lowell Observatory LOIS\cite{Taylor2000}/LOUI systems which handle
the current instrument suit on the DCT. This software system is a mature well defined control system that already has interfaces to the TCS, Guider, and AOS subsystems. If we decide that the ARC controller is the best match for NEWS
then most of the interfacing to the detector has already been done for the NIHTS instrument, an H2RG NIR instrument and Mimir, an Aladdin III instrument.

\begin{figure}
	\begin{center}
	\includegraphics[width=5in]{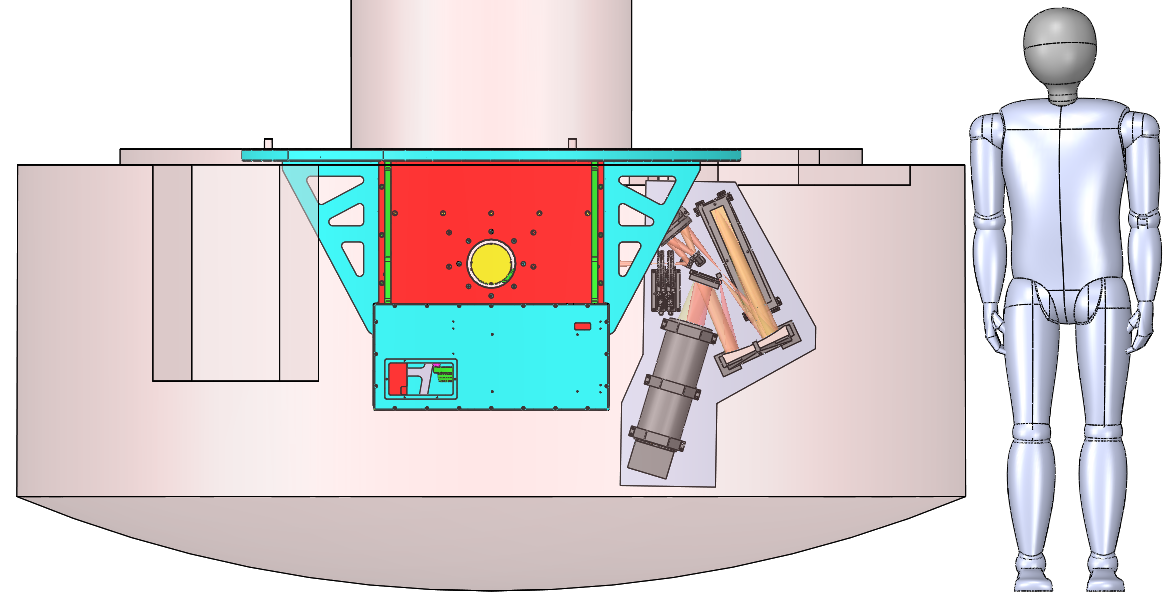}
	\end{center}
	\caption[example] 
	{\label{fig:ondct} 
	CAD rendering of of the NEWS mechanical design as it would be positioned on the DCT instrument cube. A cryostat containing the instrument would be mounted directly to one of the large ports on the instrument cube. The interference-free volume for the instrument cube is shown in transparent red. NEWS fits within the size constraints of mounting to the DCT. A six-foot-tall human figure is shown for scale.}
\end{figure}

\subsection{Thermal Management}
\label{therm}
Thermal modeling of NEWS shows that the predictive radiative load will be around 63 W at 300 K, assuming a polished aluminum cold shield wrapped in MLI and based on the current surface area of the instrument cold shield. Average high temperatures are 301 K for the hottest month of the year in Happy Jack, Arizona where DCT is located. This load can be easily handled by a CTI 1050 single stage cryodyne refrigeration unit which is capable of delivering 80 W of cooling at 78 K with enough overhead to handle any parasitic load from the wiring and the intermittent load of the motors. We will utilize thick copper strapping to the cold bench and shield with a much small copper line to the detector for the conductive paths for heat transfer. This line to the detector will have a small sapphire disk spacer to minimize the electrical noise transferred to the detector.  We will monitor and control the instrument stability, cool down and warm up rates via a Lakeshore model 331 temperature controller.

\section{Summary}
We have designed a high-resolution NIR spectrograph for the 4.3-meter Discovery Channel Telescope. NEWS achieves a resolution of $R$ = 60,000 over the full 0.8--2.5 $\mu$m range. Our design offers continuous coverage within each of its five observing modes corresponding to the photometric z-, Y-, J-, H-, and K-bands. If built, NEWS will be uniquely capable of measuring the composition and ages of field M dwarfs, including those who host planets detected by \textit{TESS}.

\acknowledgments      
 
Support for this work was provided by the Department of Astronomy and the Institute for Astrophysical Research at Boston University.  This research made use of the Massachusetts Green High Performance Computing Center in Holyoke, MA. 

\bibliography{report}
\bibliographystyle{spiebib}

\end{document}